\documentclass[draftcls, 11pt, onecolumn]{IEEEtran}
\IEEEoverridecommandlockouts
\usepackage{amsmath,amsthm,amssymb,paralist,subfigure,graphicx,algorithm2e}

\newtheorem{defn}{Definition}

\def\mc{\mathcal}

\begin{document}
	
	\title{On the Controllability of Clustered Scale-Free Networks}
	
	\author{Mohammadreza Doostmohammadian$^\dagger$,  Usman A. Khan$^\ast$
		
		\thanks{
			$^\dagger$ Mechanical Engineering Department, Semnan University, Semnan, Iran \texttt{doost@semnan.ac.ir}.

			$^\ast$ Electrical and Computer Engineering Department, Tufts University, Medford, USA \texttt{khan@ece.tufts.edu}.}}
	\maketitle

	\begin{abstract}
		In this paper, we compare the number of unmatched nodes and the size of dilations in two main random network models, the Scale-Free and Clustered Scale-Free networks. The number of unmatched nodes determines the necessary number of control inputs and is known to be a measure for network controllability, while the size of dilation is a measure of controllability recovery in case of control input failure. Our results show that clustered version of Scale-Free networks require fewer control inputs for controllability. Further, the average size of dilations is smaller in clustered Scale-Free networks, implying that potentially fewer options for controllability recovery are available.
		
	\end{abstract}

	\section{Introduction}
	Controllability and observability of complex networks have gained significant attention in the  literature \cite{Controllability_JCN,Liu_nature,lin,liu2016tutorial,jstsp14,jstsp,globalsip14,doostmohammadian2017observational,acc13}.
	Network controllability is known to be related to the concepts of \textit{dilation} and \textit{matching} in networks \cite{Liu_nature,liu2016tutorial}. A matching is a component in the network defining the \textit{structural rank} of its adjacency matrix \cite{globalsip14,doostmohammadian2017observational}. Simply, more unmatched nodes in the network implies greater rank-deficiency of the adjacency matrix. On the other hand, dilations represent the components in the network in which less number of nodes are linked (dilated) to more other nodes.  It is known that for controllability every (unmatched) node in each dilation is necessary to be controlled. These controlled nodes, to which  the control input is injected, are also known as \textit{driver nodes}. Note that the nodes in a dilation are all equivalent in terms of controllability \cite{Liu_nature}. Therefore, the size of dilation  defines the possible number of driver nodes to recover for loss/failure of a control input. This simply implies that larger dilations provide more options for controllability recovery.
	
	Controllability of complex networks \cite{cremonini2017controllability,Liu_nature}  is the topic of this paper. One well-known random model for complex networks is introduced by Barabasi and Albert \cite{barabasi_albert1999}, referred to as the Scale-Free (SF) network model.  In \cite{barabasi_albert1999} it is shown that the degree distribution of SF networks follows a power-law distribution as in real-world networks. However, the SF networks have low clustering coefficient\footnote{The clustering coefficient is defined as the fraction of neighbors of each node that are also neighbor of each other. In other words, the clustering coefficient counts the number of closed triplets (triangles) to the total number of triplets in the network. Mathematically,
		$CC = 3\frac{tr}{trp}$
		where $tr$ counts the number of triangles and $trp$ counts the number of connected triplets.}. In order to alter this issue, recently a new modified model for Scale-Free networks is proposed \cite{klemm2002highly,Holme2002clusteringScaleFree,Toivonen2006social}, based on triad formation. The new model, known as Clustered Scale-Free (CSF), is constructed based on the fact that there is high probability that two
	neighbors of one node in the complex network are connected themselves, resulting in high clustering coefficient. This property along with having small average length of shortest path
	between two nodes is sometimes referred to as \textit{small world} property \cite{ebel2002scale,wasserman1994social}.
	
	In this work, we compare the controllability of SF and CSF models. Note that the only different factor between SF and CSF networks is their  clustering coefficient (refer to \cite{Holme2002clusteringScaleFree,Toivonen2006social} for evidence of this claim). In this direction, we compare the number of unmatched nodes and average size of dilations in two types of  networks to investigate the effect of clustering coefficient. This is important because the clustering coefficient of synthetic networks is known to be tunable and algorithms are introduced in the literature to change the clustering coefficient of networks \cite{Holme2002clusteringScaleFree,serrano2005tuning,kashyap2017mechanisms,fazli2015using}. 
	We further increased the clustering coefficient in a real-world Scale-Free network  by adding more random links for closed triplet formation. The number of unmatched nodes (driver nodes) and the average dilation size is investigated, showing dependency on the change in the clustering coefficient. 
	Therefore, the results of this paper are significant as by tuning the clustering coefficient one can manage the controllability properties of synthetic complex networks. The results are specifically stated for Scale-Free types of networks which are prevalent in many real-world applications \cite{newman2003structure,faloutsos1999power}.
	
	The rest of the paper is organized as follows. In Section~\ref{sec_dil}, we introduce the concepts of maximum matching, unmatched nodes, and dilations as main factors in network controllability. In Section~\ref{sec_SF}, we discuss two main models for random networks, the SF and CSF models. We state our main results on the controllability comparison of these two models and further the effect of tuning the clustering coefficient of a real-world network in Section~\ref{sec_res}. Finally, Section~\ref{sec_con} concludes the paper.
	
	\section{Matching and Dilation: Definition and Algorithm} \label{sec_dil}
	In this section, we define the concepts of unmatched nodes and graph dilation along with some preliminary  graph notions. Next,  a polynomial-order algorithm is provided to find the maximum matching and dilations in a network.
	
	Consider the  graph $\mc{G} = (\mc{V},\mc{E})$ with $\mc{V}$ as the set of nodes and $\mc{E}$ as the set of links. Define a bipartite graph~$\Gamma=(\mc{V}^+,\mc{V}^-,\mc{E}_\Gamma)$, associated to $\mc{G}$, as a graph with two disjoint set of nodes denoted by~$\mc{V}^+$ and~$\mc{V}^-$ and the set of links denoted by~$\mc{E}_\Gamma$. Every link in~$\mc{E}_\Gamma$  starts in~$\mc{V}^+$ and ends in~$\mc{V}^-$. We have, $\mc{V}^+=\mc{V}$, $\mc{V}^- = \mc{V}$, and the link set~$\mc{E}_{\Gamma}$ is the collection of~$\{(\mc{V}_j^-,\mc{V}_i^+)|(\mc{V}_j,\mc{V}_i) \in \mc{E}\}$. In the bipartite graph define a \textit{matching}, denoted by~$\underline{\mc{M}}$,  as the subset of links that share no begin nodes in~$\mc{V}^-$ and no end nodes in~$\mc{V}^+$. Therefore, all links in~$\underline{\mc{M}}$ are  independent and mutually disjoint. Defining the size (cardinality) of the matching as its number of links, a matching with maximum cardinality/size is called \textit{maximum matching}, denoted by~$\mc{M}$. The maximum matching, in general, is \textit{not unique}. In other words, the size of the maximum matching is equal to the \textit{structural-rank} of the adjacency matrix of the graph $\mc{G}$. The structural-rank of the adjacency matrix $\mc{A}_{\mc{G}}$ of the graph $\mc{G}$ is defined as the maximum number of non-zero entries in $\mc{A}_{\mc{G}}$ that share no rows and columns \cite{rein_book}. Each of these entries in $\mc{A}_{\mc{G}}$ represent a link in the maximum matching of the graph $\mc{G}$. 
	\begin{defn} \label{def_unmatched}
		Define the set of \textit{matched} nodes, denoted  by~$\partial \mc{M}^-$, as the nodes in~$\mc{V}^-$ incident to maximum matching~$\mc{M}$. Denote by~$\delta \mc{M}$, the set of \textit{unmatched} nodes defined as~$\delta\mc{M} = \mc{V}^- \backslash \partial \mc{M}^-~$. This simply implies that a node is matched if, in bipartite graph representation $\Gamma$, it is an ending node of a link in the maximum matching; otherwise the node is \textit{unmatched}. In other words,, in bipartite graph representation $\Gamma$, the unmatched nodes are not the ending node of any link in the maximum matching $\mc{M}$. 
	\end{defn}
	
	Given the matching~$\underline{\mc{M}}$ and the bipartite graph~$\Gamma$, let define a new graph called the \textit{auxiliary graph}~$\Gamma^{\underline{\mc{M}}}=(\mc{V}^+,\mc{V}^-,\mc{E}_{\Gamma})$ as follows: keep the direction of the links in~$\mc{E}_{\Gamma} \backslash \underline{\mc{M}}$  while reversing the direction of all links in~$\underline{\mc{M}}$. Next, consider a sequence of links called the \textit{${\underline{\mc{M}}}$-alternating path}, denoted by~$\mc{Q}_{\underline{\mc{M}}}$, associated to  the matching~$\underline{\mc{M}}$ and auxiliary graph~$\Gamma^{\underline{\mc{M}}}$, as sequence of links alternating between matched links~$\mc{M}$ and unmatched links~$\mc{E}_{\Gamma} \backslash \underline{\mc{M}}$. Start the sequence with an unmatched link in~$\mc{E}_{\Gamma} \backslash \underline{\mc{M}}$ from a node in~$\delta \underline{\mc{M}}$ and every second link in~$\underline{\mc{M}}$. Further, define an \textit{${\underline{\mc{M}}}$-augmenting path}, denoted by~$\mc{P}_{\underline{\mc{M}}}$, as an alternating path starting and ending in~$\delta \underline{\mc{M}}$. Having defined these graph notions we are ready to introduce the concept of dilation as follows:
	
	\begin{defn}\label{def_dilation}
		For a maximum matching $\mc{M}$ take  every  node~$\mc{V}_i \in \delta \mc{M}$ and find the  set of  nodes in $\mc{V}^-$ in auxiliary graph~$\Gamma^\mc{M}$ that can be reached by alternating paths~$\mc{Q}_{\mc{M}}$ from~$\mc{V}_i$. This set is called a dilation~$\mc{D}$. In other words, in a dilation there is a subset $\mc{F} \subset \mc{V}$ such that $|\mc{N}(\mc{F})|<|\mc{F}|$, where $| . |$ is the cardinality of the set and $\mc{N}(\mc{F})$ represents the set of neighbors of the nodes in $\mc{F}$ defined as $\mc{N}(\mc{F})=\{\mc{V}_j|(\mc{V}_j,\mc{V}_i) \in \mc{E}, \mc{V}_i \in \mc{F} \}$. Rouphly speaking, in the graph $\mc{G}$ the links associated with a dilation represent a component in whcih less nodes point (link) to more other nodes \cite{Liu_nature}.
	\end{defn}
	
	We refer interested readers to \cite{murota} for more information regarding the graph-theoretic concepts described above. The process of finding maximum matching and dilations in a given network is summarized in the following algorithm.
	
%
	
	\begin{algorithm} \label{alg_dil}
		\textbf{Given:} System digraph $\mc{G}_A$ 
		
		Make $\Gamma$\;
		Find a matching $\underline{\mc{M}}$ \;
		Make $\Gamma^{\underline{\mc{M}}}$ \;
		\While{$\mc{P}_{\mc{M}}$ exist}{
			\For{unmatched nodes in $\delta \underline{\mc{M}}$}{  
				Find $\mc{P}_{\{\mc{M}}$ \;
				$\underline{\mc{M}} = \underline{\mc{M}} \oplus \mc{P}_{\underline{\mc{M}}}$ \;
			}
		}
		Make $\Gamma^{\mc{M}}$ \;
		\For{unmatched nodes in $\delta \mc{M}$}{  
			Find $\mc{Q}_{\mc{M}}$ in $\Gamma^{\mc{M}}$ \;
			Define $\mc{D}_i$ as nodes in $\mc{V}^-$ reachable by $\mc{Q}_{\mc{M}}$\;
		}
		\textbf{Return} $\mc{D}_i, i = \{1,...,l\}$\;
		
		\caption{Finding the maximum matching and graph dilations}
	\end{algorithm}
	
	Note that $\oplus$ in the above algorithm is the XOR operator. The first loop of the Algorithm~\ref{alg_dil} starts with a matching $\underline{\mc{M}}$ and finds a maximum matching $\mc{M}$. The second loop uses this maximum matching to find the dilations in the network $\mc{G}$. We use this algorithm to find the unmatched nodes and dilations in different types of random complex networks defined in the next section.

\begin{figure}
		\centering
		\includegraphics[width=3.5in]{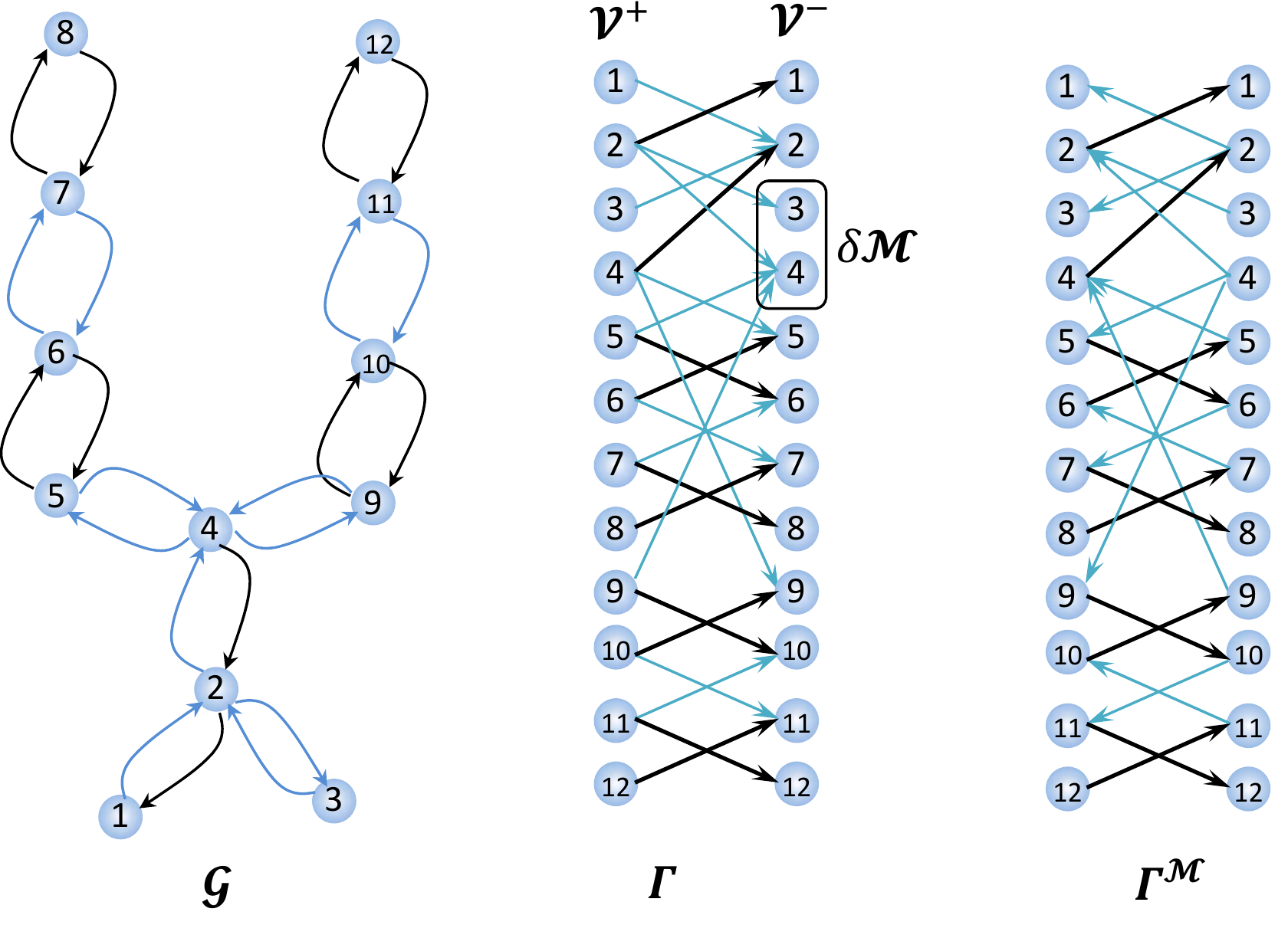}
		\caption{Example graph $\mc{G}$ with $12$ nodes is represented, illustrating the concepts of the bipartite graph~$\Gamma$, a maximum matching~$\mc{M}$, unmatched nodes $\delta \mc{M}$, and the auxiliary graph $\Gamma^\mc{M}$. The matched links are shown in black.}
		\label{fig_12nodes}
	\end{figure}

\begin{figure}
		\centering
		\includegraphics[width=4.5in]{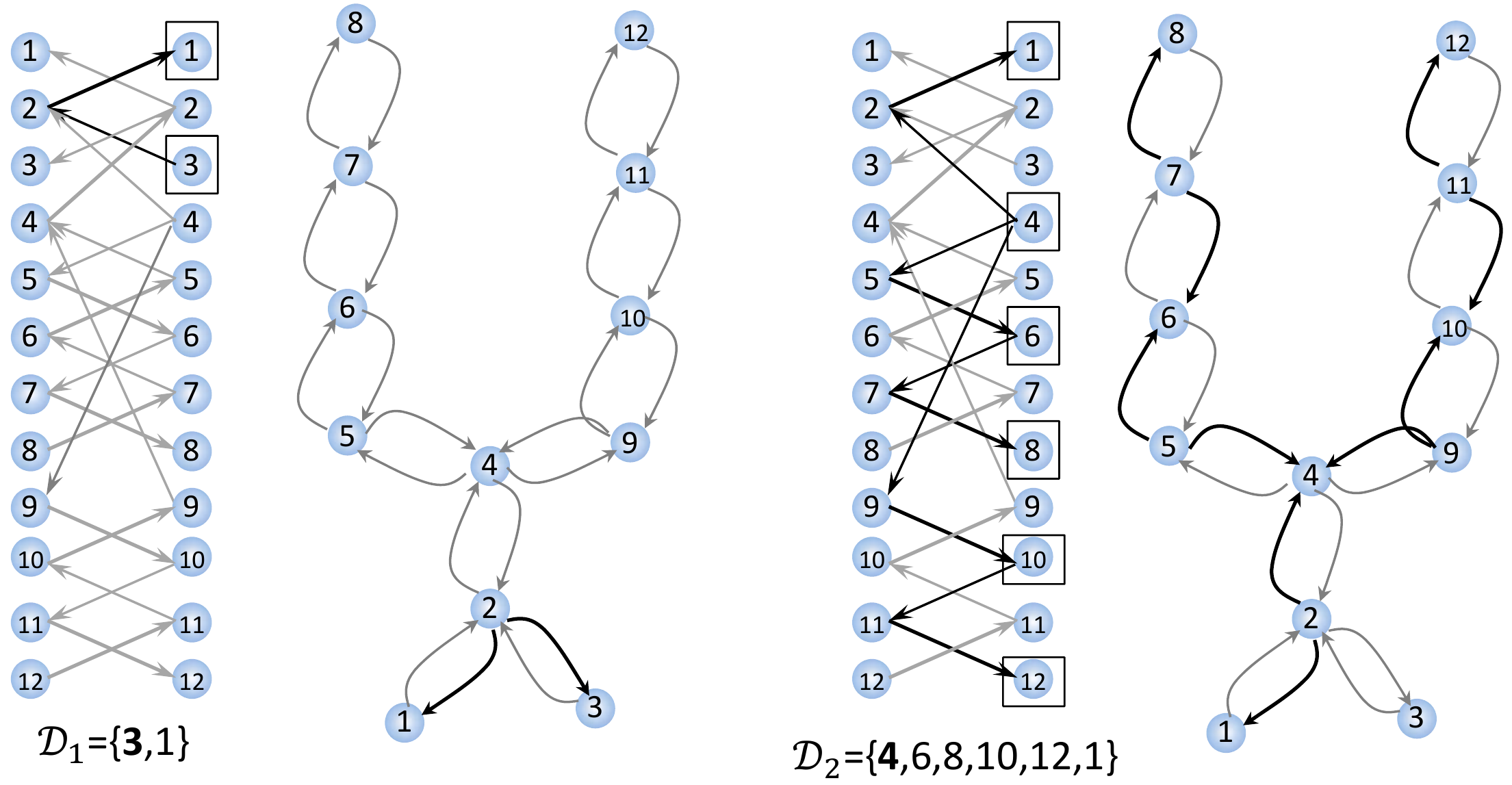}
		\caption{The procedure of finding the dilations in the graph example of Fig.~\ref{fig_12nodes} is shown. As illustrated,  all the nodes in the auxiliary graph $\Gamma^\mc{M}$ reachable by alternating paths $\mc{Q}_{\mc{M}}$ from an unmatched node represent a dilation set. The black links represent the alternating paths in the auxiliary graph $\Gamma^\mc{M}$ and the links associated with a dilation in the graph $\mc{G}$. As it can be seen the number of dilation neighbors is less than the size of dilation itself.}
		\label{fig_12nodes_dilation}
	\end{figure}

	\textit{Illustrative Example:} We provide an example graph $\mc{G}$ with $12$ nodes in Fig.~\ref{fig_12nodes} to illustrate the graph-theoretic concepts described above. 
	The structural-rank of this graph is $10$ and the size of maximum matching $\mc{M}$ is $10$. Therefore, the number of unmatched nodes in the graph is $2$. 
	In Fig.~\ref{fig_12nodes}, the links associated with one example maximum matching $\mc{M}$ are shown in black, both in the graph $\mc{G}$ and its bipartite representation $\Gamma$. Note that in the bipartite graph $\Gamma$ each link of the graph $\mc{G}$ is represented by a directed link from $\mc{V}^+$ to $\mc{V}^-$, with $\mc{V}^+$ and $\mc{V}^-$ having the same set of nodes as in graph $\mc{G}$. Following the Definition~\ref{def_unmatched}, the set of unmatched nodes in $\Gamma$ are highlighted by the black square as $\delta \mc{M} = \{3,4\}$. As it can be seen from the figure the nodes $\{3,4\}$ are not the ending node of any link in the maximum matching $\mc{M}$. $\Gamma^{\mc{M}}$ in the figure represents the auxiliary graph made by reversing the direction of all the links in $\Gamma$ except the links in the maximum matching $\mc{M}$. The auxiliary graph is used to find the dilations as shown in Fig.~\ref{fig_12nodes_dilation}. 
	The $\mc{M}$-alternating path starting at unmatched node $3$ and all $\mc{M}$-alternating paths starting at unmatched node $4$ are represented in black. As it can be seen, the links in $\mc{M}$-alternating path $\mc{Q}_{\mc{M}}$ alternate between matched links in $\mc{M}$ and unmatched links in $\mc{E}_{\Gamma} \backslash \mc{M}$. In $\Gamma^{\mc{M}}$, all the nodes reachable by the alternating paths $\mc{Q}_{\mc{M}}$ are highlighted by black squares. According to Definition~\ref{def_dilation}, these nodes represent the dilations in the graph as $\mc{D}_1 = \{3,1\}$ and $\mc{D}_2 = \{4,6,8,10,12,1\}$.  We remind the reader that all the links not included in the maximum matching $\mc{M}$ are reversed in the auxiliary graph $\Gamma^{\mc{M}}$.

	\section{SF vs. CSF Network Models} \label{sec_SF}
	One preliminary  descriptive model for complex networks  is Scale-Free (SF) model. The main characteristic of this model is power-law distribution of node degrees, which resembles the degree distribution of most real  networks, including social networks, technological networks, Internet, economic networks, etc \cite{newman2003structure,faloutsos1999power}. The most well-known construction procedure of the SF network is proposed by Barabasi and Albert \cite{barabasi_albert1999}. This recursive procedure starts with a small \textit{initial  seed} graph composed of few nodes (simply consider, for example, a line graph of few nodes). At each iteration, a new node is added to the network making new random connections with the old nodes. The probability that the new node make connection to the old nodes is proportional to the node degree. Simply, the new node prefers to connect to high degree nodes, and thus, implying the name \textit{preferential attachment} method.
	
\begin{figure}[hbpt!]
		\centering
		\includegraphics[width=2in]{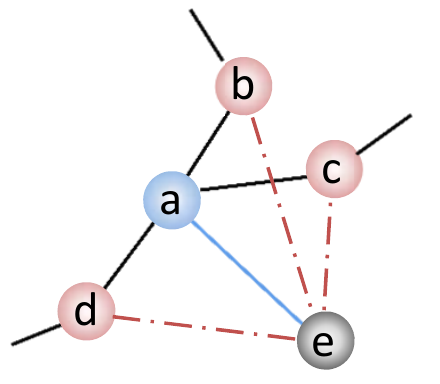}
		
		\caption{The triad formation in CSF networks is illustrated. The new node '$e$' makes connection to the preferentially attached node '$a$' in the network. The dashed lines are possible options for node '$e$' to connect with one of the neighbors  of the node '$a$'. Making a link with a neighbor node '$b$', '$c$', or '$d$' forms a triad. The dashed links show the \textit{possible} options for making a triad. The triad formation is stochastic as the node '$e$' \textit{randomly} connects to one (or more) of the neighboring nodes '$b$', '$c$', or '$d$' based on the preferential attachment.  The triad formation increases the clustering-coefficient of the CSF networks over the SF networks.}
		\label{fig_triad}
	\end{figure}

	The clustering-coefficient of the networks made based on the preferential attachment procedure is low, while in contrast, many real-world networks (including social networks) are known to have high clustering-coefficient \cite{Toivonen2006social,ebel2002scale,liljeros2001web}. Therefore the concept of
	clustered networks is introduced in the literature \cite{klemm2002highly,Holme2002clusteringScaleFree,Toivonen2006social}. The most well-known model for such networks is Clustered Scale-Free (CSF) model, based on \textit{triad formation} \cite{Holme2002clusteringScaleFree,Toivonen2006social}. The network growth procedure for such networks is similar to the preferential attachment. First, an initial seed graph is considered. Then, the new node is added to the network making $m_r$ connections to the preferentially attached nodes. But, further, the new node makes $m_s$ random connections to the neighbors of the preferentially attached nodes, see Fig.~\ref{fig_triad}. This method increases the number of triads in the network and, therefore, results in higher clustering-coefficient. note that the triad formation method is closely related to the definition of the clustering coefficient. Real-world examples of such clustered networks can be found in \cite{ebel2002scale,liljeros2001web,eguiluz2002epidemic}.
	
	It should be noted that the procedures for constructing both SF and CSF networks are \textit{stochastic} and not deterministic. This is because both networks are based on the preferential attachment method. Particularly, for triad formation in CSF networks the new node makes $m_s$ links to the neighbors of the other node based on the preferential attachment. In other words, the new node \textit{randomly} connects to one or few neighbors while the probability of the connection is proportional to the neighbor's degree. In this method, it is more likely that the neighboring node with higher degree makes a triad with the new node, see \cite{Holme2002clusteringScaleFree,Toivonen2006social} for more details.
	
	\section{Main Results} \label{sec_res}
	Recall that, the controllability of networks is to great extent related to the number of unmatched nodes and the size of dilations. More unmatched nodes in the network require more control inputs to derive the network towards desired state. On the other hand, the size of dilation in the network indicates the possible options to recover for loss of controllability. If the control input to an unmatched node fails, injecting proper control input to other nodes in the same dilation may recover the controllability\footnote{We put the topic of control-input recovery for future research direction.}. In this direction, we first compare the number of unmatched nodes and average size of dilation in two main random models, the SF and CSF networks.
	
	The networks considered for simulations range from $100$ nodes to $1000$ nodes. Each SF network is constructed based on the preferential attachment method, where at each iteration the new node makes $2$ new links with the old nodes.\footnote{We assume 2 new connections without loss of generality. Any number of new links may be considered for preferential attachment and triad formation. The main point is that the total  number of new connections in both SF and CSF networks must be the same. This is because the average node degree and number of links must be similar in both networks for the sake of comparison.} For CSF networks, $1$ link is considered for preferential attachment, and $1$ link for triad formation. Therefore, the total number of links and the average node degrees are similar in both SF and CSF networks of the same size.
	
\begin{figure}
		\centering
		\includegraphics[width=3in]{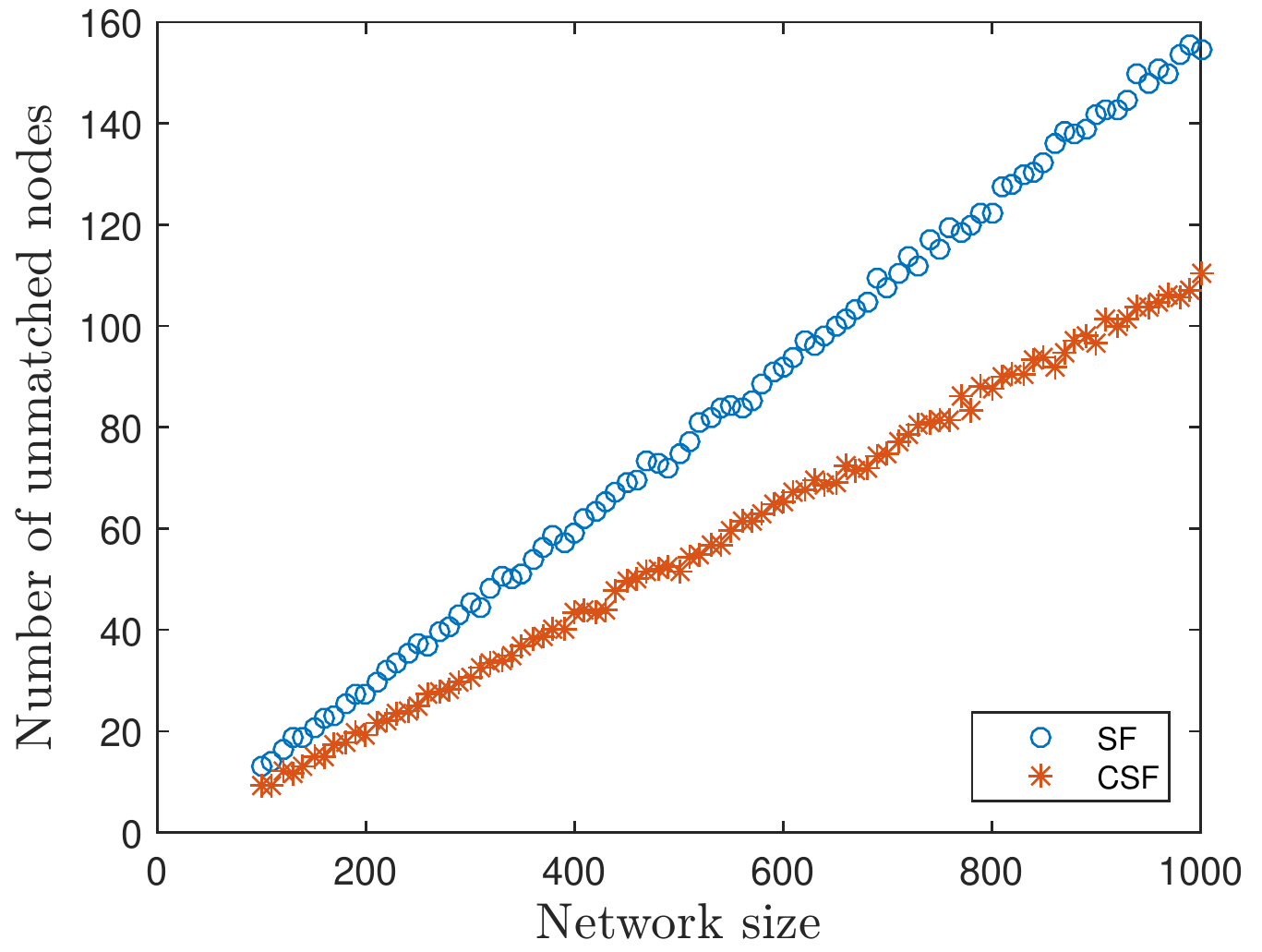}
		
		\caption{The number of unmatched nodes versus network size for SF and CSF network types are compared. The SF networks contain more number of unmatched nodes as compared to CSF networks. }
		\label{fig_unmatched}
	\end{figure}

	The number of unmatched nodes in SF and CSF networks are determined using the first part of Algorithm~\ref{alg_dil}. The results are shown in Fig.~\ref{fig_unmatched}.
	For each point in the figure we performed a Monte-Carlo simulation and the number of unmatched nodes is averaged over $100$ realizations of networks with the same size. As it can be seen, the number of unmatched nodes in SF networks is more than CSF networks. This implies that for controllability of clustered model of Scale-Free networks fewer control inputs (to be injected into driver nodes) are required.
	
\begin{figure}
		\centering
		\includegraphics[width=3in]{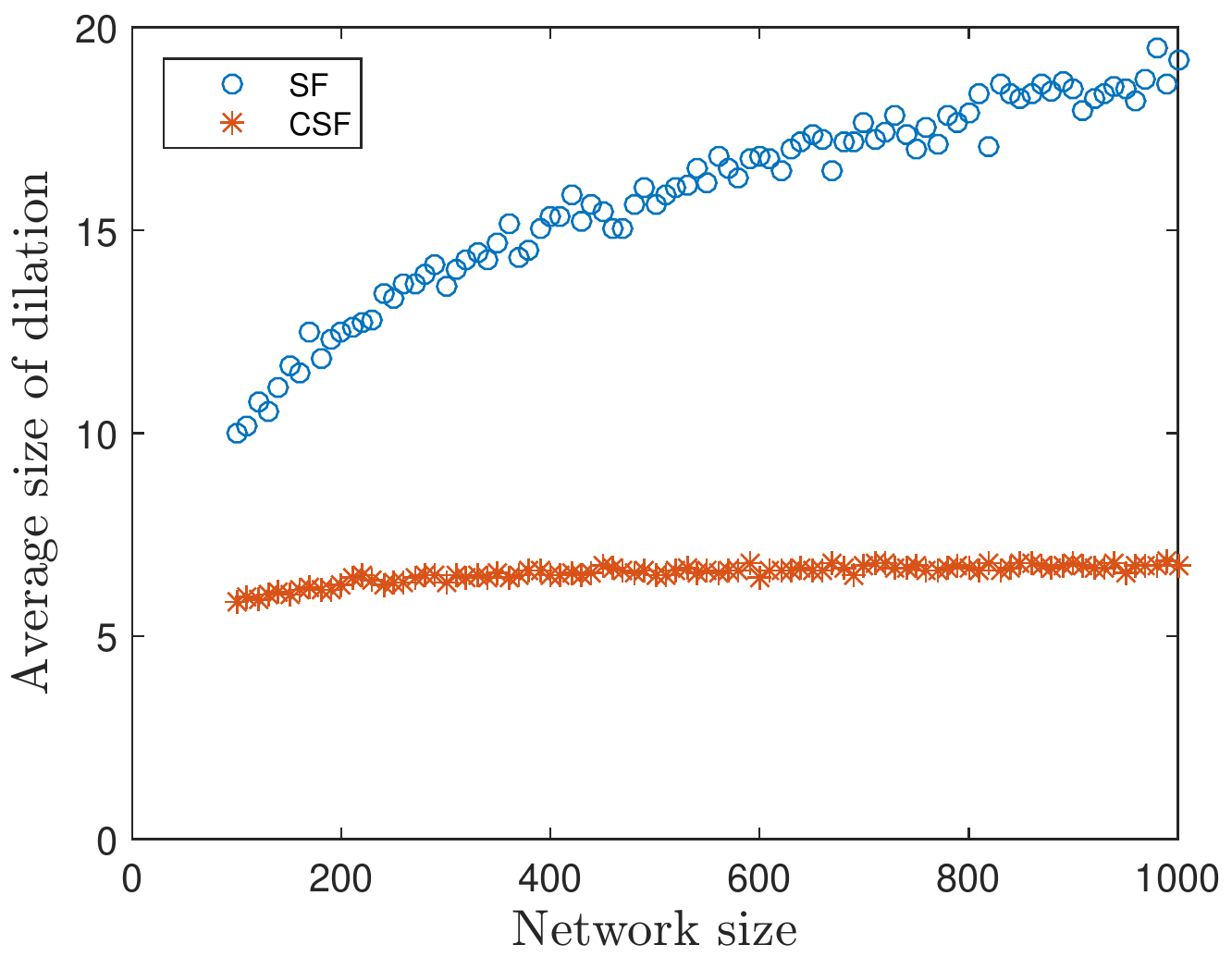}
		\caption{The average size of dilations versus network size for SF and CSF network types is shown. The SF networks contain larger dilations, in average, as compared to CSF networks. }
		\label{fig_dilation}
	\end{figure}

	Next, using Algorithm~\ref{alg_dil}, we find the dilations in the SF and CSF networks. For each network size, we find the average size of dilations for $100$ realizations of networks. The result is shown in Fig.~\ref{fig_dilation}. As it can be seen, the average size of dilations in SF networks is greater than CSF networks. This implies that, in case of control failure, for clustered version of Scale-Free networks there are fewer options to recover the loss of controllability. Also, note that by increasing the size of the network the average size of dilation in SF networks increases, while in CSF networks the average dilation size is less dependent on the network size.
	
\begin{figure}
		\centering
		\includegraphics[width=3in]{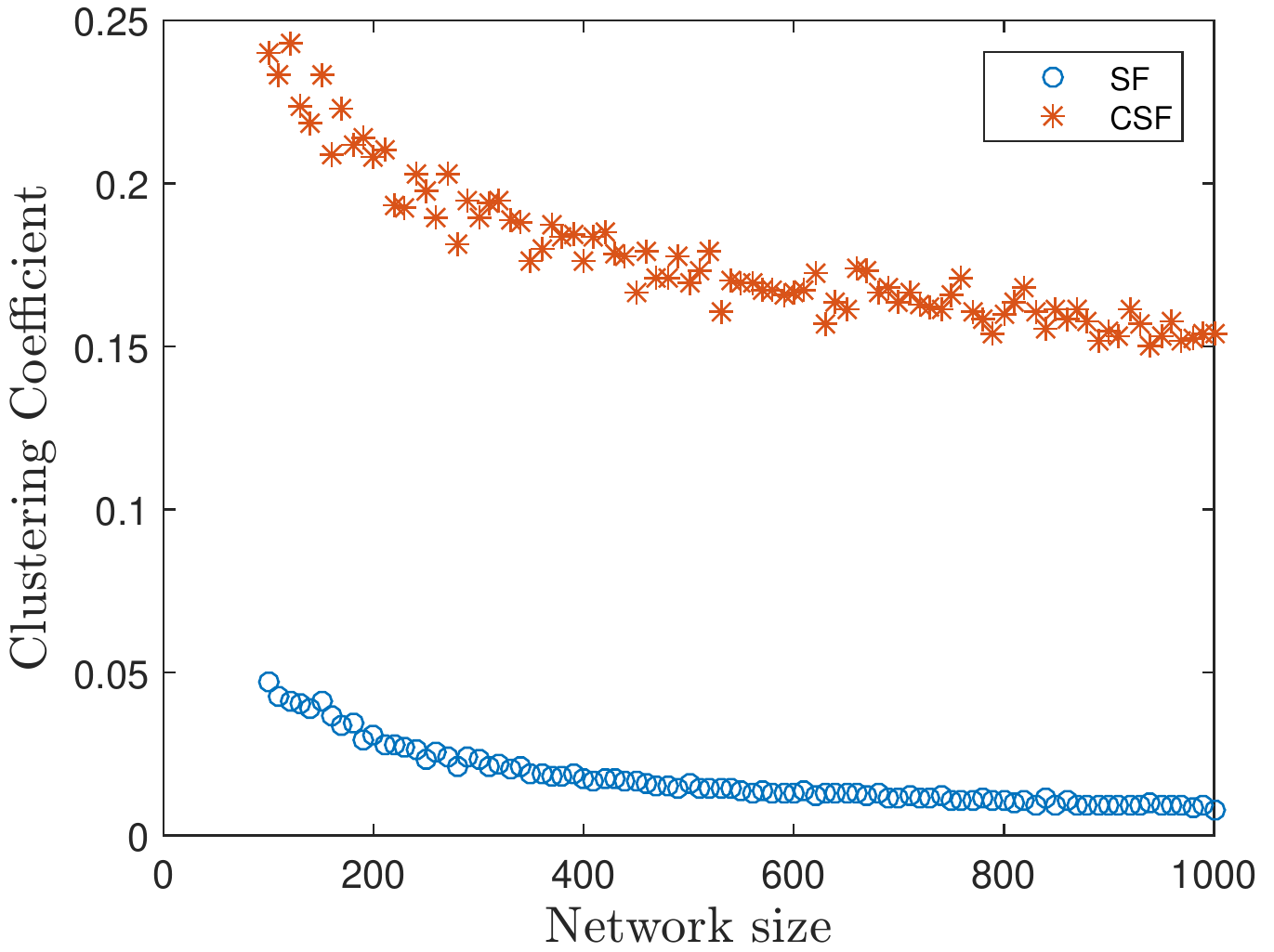}
		\caption{The clustering coefficient of SF and CSF networks versus network size are compared. As expected, the SF networks have lower clustering coefficient as compared to CSF networks. }
		\label{fig_clust_CSF}
	\end{figure}

	We further compare the clustering coefficient in SF and CSF networks of these $100$ realizations in Fig.~\ref{fig_clust_CSF}.
	As it is clear the clustering coefficient is lower in SF networks as compared to CSF networks of the same size. Further, by comparing Fig.~\ref{fig_clust_CSF} with Fig.~\ref{fig_unmatched} and Fig.~\ref{fig_dilation}, we observe that by increase in the clustering coefficient the number of unmatched nodes and average size of dilations are decreased.
	
\begin{figure}
		\centering
		\includegraphics[width=3in]{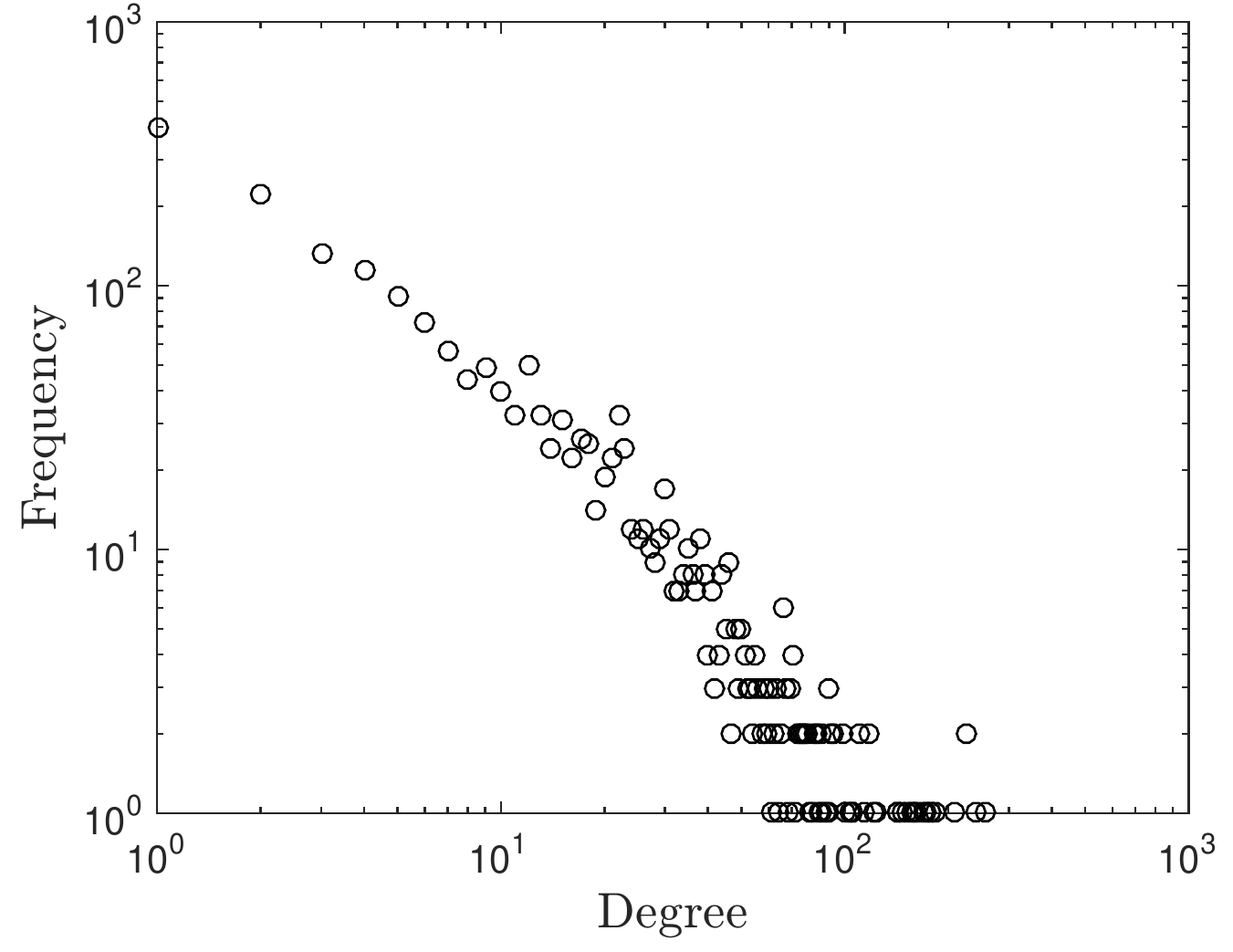}
		\caption{The degree distribution of the real network representing the online community of students in University of California, Irvine is shown. }
		\label{fig_uci}
	\end{figure}

	\textit{Real network case study:} we consider a real Scale-Free network and investigate the relation of clustering coefficient with the number of unmatched nodes and average size of dilations in this network. This network represents the interactions among users of an online community of students from the University of California, Irvine \cite{konect}. This network contains $1899$ nodes and $13838$ links, where a node represents a student user and links represent online communication among users. The degree distribution of this network is represented in Fig.~\ref{fig_uci}, which shows the Scale-Free property.
	
\begin{figure}
		\centering
		\includegraphics[width=3in]{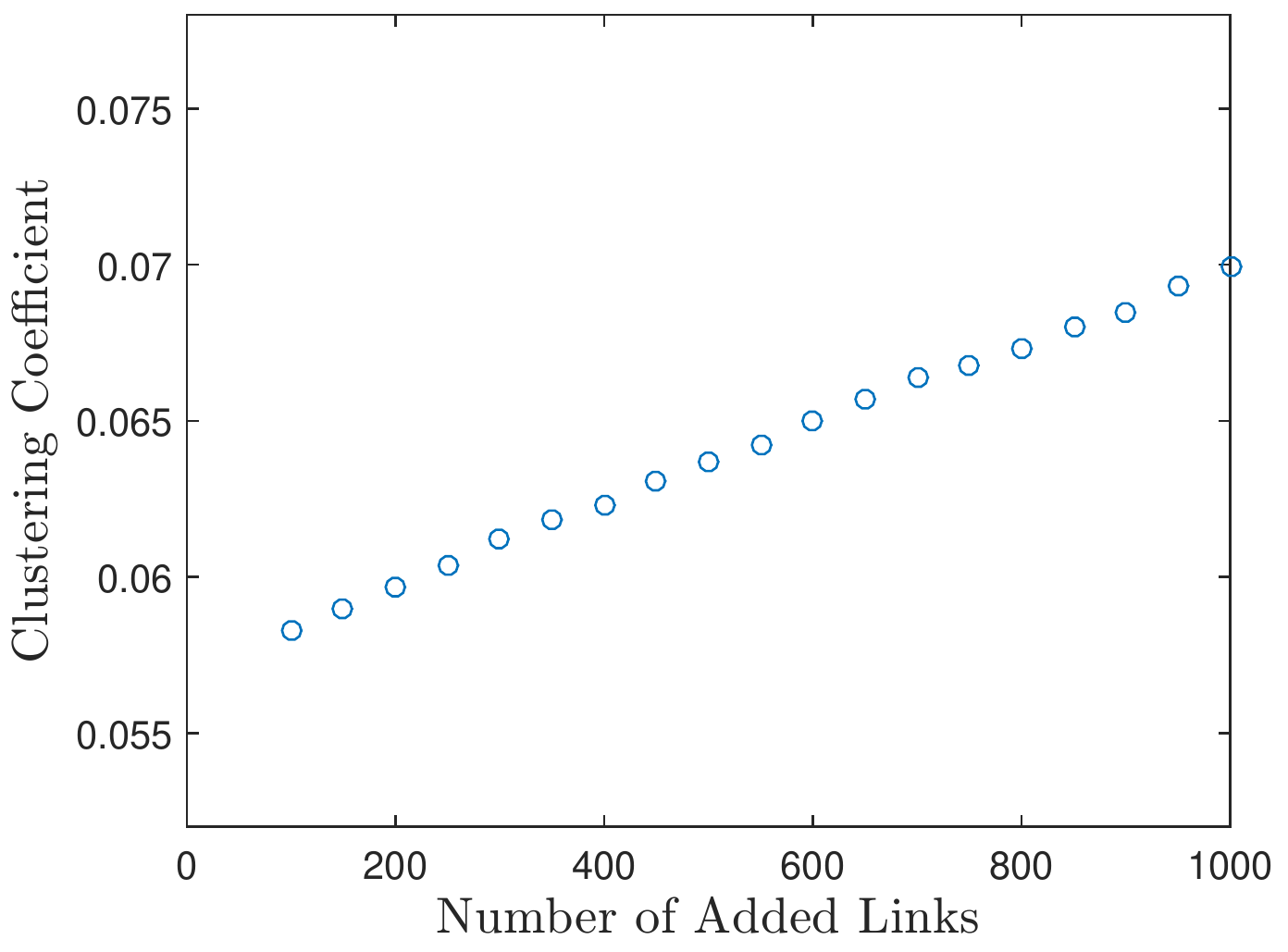}
		\caption{The change in the clustering coefficient vs. the number of added links (for closed triplet formation) in the network of Fig.~\ref{fig_uci} is shown. By increasing the number of closed triplets the clustering coefficient is increased.}
		\label{fig_clust}
	\end{figure}

\begin{figure}
		\centering
		\includegraphics[width=3in]{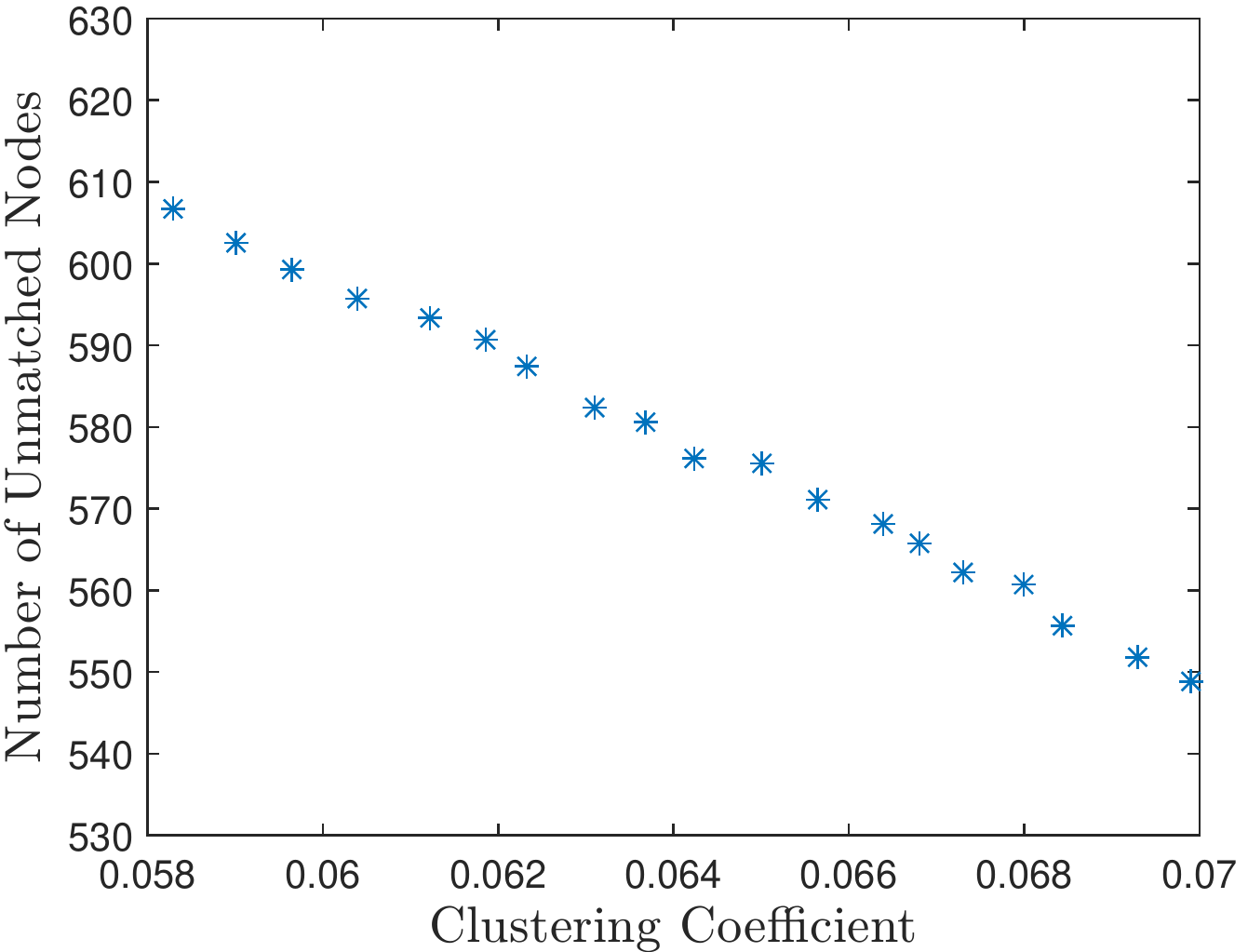}
		\caption{The change in the number of unmatched nodes vs. the clustering coefficient in the network of Fig.~\ref{fig_uci} is shown. The number of unmatched nodes is decreased by increase in the clustering coefficient.}
		\label{fig_uci_unmatched}
	\end{figure}

	We analyze the effect of change in clustering coefficient on the number of unmatched nodes for this network. Based on the definition of the clustering coefficient, we directly increase the number of closed triplets (or triads) in the network as follows:\footnote{It should be noted, although we consider an online social network as an example Scale-Free network, any real-world synthetic network may be considered for triad link addition to increase the clustering coefficient. This work is not restricted to social networks, but general industrial and technological networks where the link addition and concepts of control theory are more applicable and achievable. } two nodes are randomly chosen and if they share a neighboring node then they are directly connected via a link. The probability of choosing a node is proportional to its degree. This is to preserve the power-law degree distribution and Scale-Free property in the network. This method increases the number of closed triplets in the network. The change in the clustering coefficient is shown in Fig.~\ref{fig_clust}. Each point in the figure is averaged over $10$ realizations. 
	As it can be seen, by increasing the number of closed triplets (or triads) the clustering coefficient is increased. We check the change in the number of unmatched nodes in the same realizations of network and the results are shown in Fig.~\ref{fig_uci_unmatched}.
	It is clear that the number of unmatched nodes is decreased by increase in the clustering coefficient (and the number of triads) in the network.
	
\begin{figure}
		\centering
		\includegraphics[width=3in]{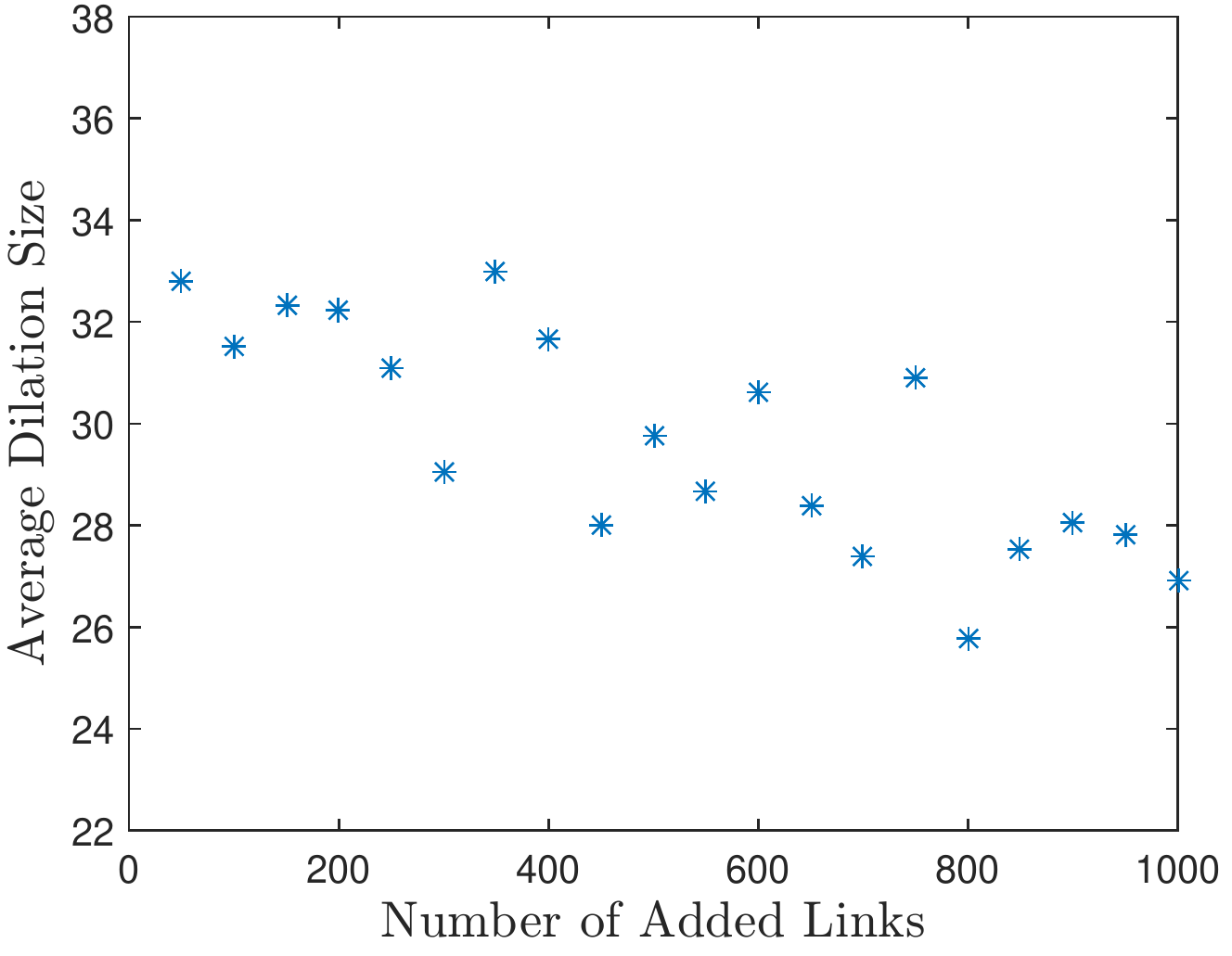}
		\caption{The change in the average size of dilations vs. the added number of links as closed triplets in the network of Fig.~\ref{fig_uci} is shown.}
		\label{fig_uci_dilation}
	\end{figure}

	Regarding the size of dilations, it should be noted that it depends both on the clustering coefficient and the number of links in the network. In general, adding more links increases the size of dilations in the network. On the other hand, more number of closed triplets and higher clustering coefficient reduces the average size of dilations. Therefore, there is a trade-off between the added number of links and effect of clustering coefficient. This can be seen in Fig.~\ref{fig_uci_dilation}. In this figure, each point is averaged over $10$ realizations.
	It can be seen from the figure that in general the average size of dilations is decreased under the effect of clustering coefficient. However, because of random nature of procedure for adding links and the general increase in the average size of dilation by increasing the number of links, the trend is not uniform. This is because, in one hand, the size of dilations is inversely dependent to the clustering coefficient, while, on the other hand, it proportionally dependent to the number of added links; however, the effect of high clustering dominates the increase in the number of links.  Note that, this is not an issue for the simulations in Fig.~\ref{fig_dilation}, because in that case the number of links in both SF and CSF networks are the same while the clustering coefficient is different. In fact, because of adopting the triad formation method, the number of closed triplets is increased in CSF networks while the total number of links is similar in both SF and CSF networks. Therefore, based on Fig.~\ref{fig_dilation}, the effect of clustering coefficient on the average dilation size can be directly deduced.
	
	\section{Concluding Remarks} \label{sec_con}
	Based on the results of previous section, we observe fewer unmatched nodes in CSF networks \textit{of the same size} as SF networks. This implies that CSF networks require fewer  control inputs (as compared to SF networks) to derive the network towards the desired state. 	
	On the other hand, the average size of dilations are smaller in CSF networks \textit{of the same size} as compared to SF networks. This implies that in case of failure/loss of a control input there are fewer options of driver nodes to recover for the loss of controllability. Note that the SF and CSF networks are similar in terms of most graph properties including \textit{power-law degree distribution}, \textit{small average geodesic length}, \textit{existence of community structure}, \textit{assortative mixing} and their only difference is the \textit{clustering coefficient} \cite{Holme2002clusteringScaleFree,Toivonen2006social}. This is because both networks are constructed based on the preferential attachment method. Note that the CSF procedure, similar to SF procedure, preserves the power-law degree distribution and keeps many other network characteristics unchanged (except the clustering coefficient). Random rewiring of the links in the SF network may not necessarily result in a power-law degree distribution and, for example, the increase in the number of random links   may result in a network similar to \textit{Erdos-Renyi} model \cite{van2010spectral}. Further, since the rewiring is random, it is not necessarily result in increase (or decrease) in the clustering coefficient. Therefore, we cannot compare the effect of clustering coefficient by random rewiring of the links (using  Monte-Carlo simulation) as, for example, the degree distribution may change among other properties. In our simulations, comparing the same size networks in Fig.~\ref{fig_unmatched}, Fig.~\ref{fig_dilation}, and Fig.~\ref{fig_clust_CSF} implies that the clustering coefficient is a key factor in the controllability of Scale-Free networks. 
	To further verify these results we investigate the link addition based on closed triplet (triad) formation in the real Scale-Free network under study. By randomly increasing the number of closed triplets (and increasing the clustering coefficient) while preserving the degree distribution, the number of unmatched nodes in the network is reduced. As expected, this implies an inverse dependency between the clustering coefficient and the number of unmatched nodes (or driver nodes) for controllability. On the other hand, the size of dilation is dependent to both the clustering coefficient and the number of added links. Although by adding more links the average size of dilations must be increased, increase in the clustering coefficient resulting from the added number of closed triplets causes reduction in the average size of network dilations. This is more clear in the results of Fig.~\ref{fig_dilation} comparing the \textit{same size} CSF and SF networks. In this figure, one can compare the networks with the same number of links, where the only difference is their clustering coefficient according to Fig.~\ref{fig_clust_CSF}. It is clear that clustered version of SF networks have smaller dilations in average.
	
	It should be mentioned, the result of this paper can be applied for controllability of different Scale-Free networks. It is known that many industrial, technological, and even economical networks structurally formed by the Scale-Free representation. Therefore, by tuning the clustering properties of such synthetic Scale-Free networks, the number of necessary control inputs (or driver nodes) and recovery of control failures can be managed, which is the direction of our future research. Note that man-made networks are prevalent in industrial applications, for example in sensor networks \cite{jstsp,kruzick2018structurally}, multi-agent systems \cite{li2014multi}, robotic networks \cite{bullo2009distributed}, Internet of Things (IoT) applications \cite{atzori2010internet}, and Cyber-Physical Systems (CPS) \cite{ISJ}. In such applications it is typical to design and engineer the network of devices, for example, for monitoring purposes. The results of this paper can be applied for design of these networks such that by increasing the clustering the number of driver nodes for controllability is reduced.   
	The results of this paper can be further extended to the dual concepts of observer nodes and network contractions for estimation recovery \cite{doostmohammadian2017recovery,doostmohammadian2017observational,icassp16}.
	We should emphasize that the main contribution of this paper is investigation of the properties of the Clustered Scale Free networks, particularly the clustering coefficient. In this paper, we do not introduce a  method for the control of real-world networks, as it is the direction of our future research.

\bibliographystyle{IEEEbib}
	\bibliography{biblio}
	
\end{document}